\documentstyle[12pt]{article}
\begin{document}
\baselineskip 15pt plus 2pt
\newcommand{\be}{\begin{equation}}
\newcommand{\ee}{\end{equation}}

\hspace{10.5cm}
BRX-TH-354
\vspace{1cm}
\begin{center}
\begin{bf}
\begin{Large}
Avoidance of Classical Singularities \\
in Quantized Gravitational Dust Systems \\
\end{Large}
\end{bf}
\vspace{1cm}
Yoav Peleg\footnote{This work is supported by the NSF grant
PHY 93-15811}\\
\vspace{0.3cm}
Physics Department, Brandeis University, Waltham, MA 02254\\
\vspace{0.2cm}
{\em Peleg@Brandeis}\\
\vspace{1.5cm}
\begin{large}
ABSTRACT\\
\end{large}
\end{center}
\vspace{0.2cm}
We use the canonical quantization of spherically
symmetric dust universes, and calculate the expectation
values of the quantized metric, $<\Psi|\hat{g}_{\mu \nu}
|\Psi>$. Though the classical solutions are singular,
and the wave functions have no zero support on singular
geometries, the expectation value (for a general quantum
state) is everywhere regular. For a quasi-classical (coherent)
state, $|\alpha>$, the expectation value, $<\alpha|\hat{g}|
\alpha>$ describes a universe (or star) that collapses to
a minimum radius, the Planck radius, and re-expands again.

\newpage

The canonical quantization of simple dust universes
is known for many years [1], and it was shown
recently that the quantum states describing those
systems are bound states [2,3].
In this work we discuss the issue of
singularities of those models.

We use the geometrical coordinates, $G=c=1$, and $\hbar \neq 1$.

Consider first a homogeneous close universe. It can be
described by the Friedmann metric with the line element
   \be
ds^{2} = - N_{\perp}^{2}(t) dt^{2} + a^{2}(t) d\Omega_{3}^{2}
   \ee
where $d\Omega_{3}$ is the volume element on the 3-sphere.
The matter in this universe is dust, with the
Lagrangian [3,4] $L_{D} = - 8\pi \int \sqrt{-g} \rho
U^{\mu} U_{\nu} d^{3}x$. Where $\rho$ is the dust density.
The total Lagrangian including
the gravitational part, can be written as [2,3]
   \be
L = L_{grav} + L_{dust} = 12\pi \int_{0}^{\pi} \mbox{sin}^{2}
\chi d \chi \left[ \frac{a \dot{a}^{2}}{N_{\perp}} -
N_{\perp}(a - a_{0}) \right]
   \ee
where $a_{0}$ is the maximum radius of the universe. The
Hamiltonian of this system is
   \be
H = N_{\perp} \left[ \frac{1}{24 \pi^{2}} \frac{P_{a}^{2}}{a}
+ 6\pi^{2} (a - a_{0}) \right]
   \ee
where $P_{a} = \partial L / \partial \dot{a}$.
The classical solution of this simple system is the
well-known cycloidal solution [5]
   \be
a(t) = \frac{1}{2} a_{0} ( 1 + \mbox{cos}(t) )
   \ee
where we chose the gauge in which $N(t)=a(t)$.
The solution (4) describes a universe, starting with a
maximum radius, $a_{0}$, that collapse to the singularity,
$a=0$.

The quantization of this system is elementary,
using the Hamiltonian (3), we can get
the Wheeler-DeWitt equation [2,3] (or the Schr\"{o}dinger
equation)
   \be
\frac{\partial \hat{H}}{\partial N_{\perp}} \psi(a) =
\left[ \frac{-\hbar^{2}}{24 \pi^{2}} \frac{1}{a}
\frac{d^{2}}{d a^{2}} + 6\pi^{2} (a - a_{0})
\right] \psi(a) = 0
   \ee
where we have used the coordinate representation
$<a|\Psi>=\psi(a) ~,~ \hat{a}=a ~,~ \hat{P}_{a} = -i\hbar
\partial / \partial a$. If we define $x = a - \frac{1}{2}a_{0}$,
we get the harmonic oscillator equation
   \be
\left[ - \frac{\hbar^{2}}{2m} \frac{d^{2}}{d x^{2}} +
\frac{1}{2} m \omega^{2} x^{2} \right] \psi(x) = E\psi(x)
   \ee
where $m = 12 \pi^{2}$, $\omega = 1$ and
$E = 3 \pi^{2} a_{0}^{2}$. We see that the quantum
state describing this universe is a bound state, $\psi_{n}(x)$,
with the spectrum
   \be
E_{n} = 3 \pi^{2} (a_{0}^{2})(n) = \hbar
\left( n + \frac{1}{2} \right)
   \ee
We see that the initial radius of the universe can take only
discrete values [2],
   \be
a_{0}(n) = \frac{1}{\sqrt{3}\pi} \sqrt{n + 1/2}
   \ee

The Oppenheimer-Snyder (OS) collapsing star is a
spherically symmetric homogeneous dust star that collapses
to form a Schwarzschild black hole [6]. The interior is
a slice of the homogeneous Friedmann universe.
One can use (1), but the angle $\chi$ is limited to
the region $[0,\chi_{0}]$.
Where $\chi_{0}$ is a constant angle smaller than $\pi / 2$.
The star's surface is at $\chi = \chi_{0}$. Outside the
star we have the (static) Schwarzschild space-time.
The matching conditions on the star's surface give
   \be
2M = a_{0} \mbox{sin}^{3}\chi_{0} ~~~~~\mbox{and}~~~~~
R_{0} = a_{0} \mbox{sin}\chi_{0}
   \ee
where $R_{0}$ is the initial Schwarzschild radius of the
collapsing star, and $M$ is its mass.

The Lagrangian in that case is $L = L_{inside} + L_{outside}$.
Where $L_{inside}$ is (2) (where the $\chi$-integration is from
zero to $\chi_{0}$ only). The outside Lagrangian describes an
empty Schwarzschild space-time, and when we study the
semiclassical states of this system, it does not contribute.
This is because $\psi_{_{WKB}} \sim \mbox{exp}(i S_{class}/\hbar)$,
and $S_{class}(\mbox{Sch.}) = 0$. The classical solution
in that case is again (4). It describes a star, initially
with a radius $R_{0}$, which collapses to the Schwarzschild
singularity, $a=0$.

The Wheeler-DeWitt equation in this case, is very similar to (5),
with the only difference that now $0 \leq \chi \leq \chi_{0}$.
The states are bound states, and
the quantization conditions are [3]
   \be
\frac{1}{4} (\alpha_{0} a_{0}^{2})(n) = \hbar \left( n
+ \frac{1}{2} \right)
   \ee
where $\alpha_{0} = 12 \pi \int_{0}^{\chi_{0}} \mbox{sin}^{2}
d \chi$. From (9) and (10) we see that the mass and initial radius
of the Oppenheimer-Snyder star are subject to a quantization [3].

One way of avoiding the classical singularity is if the
wave function has zero support on singular geometries.
Namely, if $\psi (g_{\mu \nu}=g_{\mu \nu}^{(sing)}) =0$.
In our
case the singular geometries are the ones for which $a=0$.
This is the ``big crash" in the cycloidal solution, and the
black hole singularity in the OS model.
If we had $\psi(a=0) = 0$,
then the probability of having a singularity is zero,
${\cal P}_{singular} = |\psi(a=0)|^{2} = 0$.
Unfortunately, this is not the case in our model.
The wave functions
that satisfy the Wheeler-DeWitt equation are the
standard harmonic oscillator wave functions, and those
do not vanish for $a=0$, or $x = -\frac{1}{2}a_{0}$.
Does this mean that we have singularities also in
the quantum model? In other words, is the
condition $\psi(a=0)=0$, necessary for avoiding the classical
singularity?

We will suggest a much weaker condition, based on the
assumption that the {\em basic} physical quantity is the
metric, $g_{\mu \nu}$. In that case, in the quantum
theory, if the system is in a state $|\Psi>$, one should
calculate the metric expectation value $<\Psi| \hat{g}_{
\mu \nu} |\Psi>$. This is a real matrix. The properties
of this real matrix determine the geometric structure
of the ``quantum system", especially, the singular
structure of it. Using this weaker condition, we will see
that the above models can be ``quantum mechanically
singularity free".

Excplicitly, we propose the following criteria for a
``quantum singularity": \\
{\em Given a quantum state of the system, $|\Psi>$, one should calculate
the expectation value of the metric $<\Psi| \hat{g}_{\mu \nu}
|\Psi>$. The quantum system is singular only if its expectation
value is.}

The metric can be singular for some states but not for others.
So this criteria may serve as a ``wave function selector".
The stronger criteria, $\psi(a=0)=0$, is not satisfied by
{\em any} state in the Hilbert space of the solutions of (6).
We are going to see that the weaker one,
regularity of $<\hat{g}_{\mu \nu}>$, {\em is} satisfied by a general
quantum state.

Consider first the ``most classical situation".
This is described by the quasi-classical coherent state,
$|\alpha>$, which satisfies (see the solution (4))
   \be
\hat{a}_{x} |\alpha(t)> = \alpha(t)|\alpha(t)> =
\frac{1}{2\sqrt{2}} a_{0}e^{-i \omega t} |\alpha>
   \ee
where $\hat{a}_{x} = \frac{1}{2}(\hat{X} + \hat{P}_{x})$.
Now let us calculate $<\alpha| \hat{g}_{\mu \nu} |\alpha>$.
Using (1) we see that the metric (in the gauge $N(t)=a(t)$) is
   \be
ds^{2} = - a^{2}(t) dt^{2} + a^{2}(t)d\chi^{2}
+ a^{2}(t) \mbox{sin}^{2}\chi d\Omega_{2}^{2}
   \ee
and we need to calculate only $<\alpha| \hat{a}^{2} |\alpha>$.
Where $\hat{a}$ is the operator $\hat{X} - \frac{1}{2}a_{0}$.
This is elementary
   \be
<\alpha | \hat{a}^{2} | \alpha> = \frac{1}{2} \left(
\alpha + \alpha^{*} \right)^{2} + \frac{a_{0}}{\sqrt{2}}
\left( \alpha + \alpha^{*} \right) + \frac{a_{0}^{2}}{4}
+ \frac{\hbar}{2}
   \ee
and using (11), we get
   \be
<\alpha| \hat{a}^{2} |\alpha> = \frac{a_{0}^{2}}{4}
( 1 + \mbox{cos}(t))^{2} + \frac{\hbar}{2}
   \ee
The first term on the r.h.s of (14) is exactly the classical
solution (4), and the second term is the
quantum correction. The quantum correction depends of course
on the quantization scheme that we use. Namely, it depends
on the ordering that we choose. To get (13) we use the
``symmetric ordering", namely,
$X^{2} = \hat{a}_{x}^{\dagger} \hat{a}_{x} + \hat{a}_{x}
\hat{a}_{x}^{\dagger} + \hat{a}_{x}^{2} +
(\hat{a}_{x}^{\dagger})^{2}$.
If we define the $p$-ordering such that
$X^{2} = (1-p) \hat{a}_{x}^{\dagger} \hat{a}_{x}
+ p \hat{a}_{x} \hat{a}_{x}^{\dagger} + \hat{a}_{x}^{2}
+ (\hat{a}_{x}^{\dagger})^{2}$,
we see that the quantum
correction goes like $ p \hbar $. So for example if we use
the Wick order, $p=0$, we get zero quantum correction.

{}From (1) we see that a singular quantum metric is a one
in which the expectation value $<\Psi| \hat{a}^{2} |\Psi>$,
vanishes for some $t$. But
according to (14) this is impossible\footnote{Unless
$p=0$.}. The expectation value $<\alpha|\hat{g}|\alpha>$
is everywhere regular, it does not vanish for any $t$.
Remember that
$|\alpha>$ is the semiclassical (coherent) state, and so
one may suspect that the quantum corrections are at least as
strong for a general quantum state. This is indeed the case.
The Hilbert space of these models is spanned by $\{|n>\}$,
the harmonic oscillator eigenstates. For a general eigenstate,
$|n>$, we have (using the $p$-order)
   \be
<n| \hat{X}^{2} |n> = \hbar n + p\hbar
   \ee
the first term on the r.h.s of (15),
$\hbar n$, is the one that contributes to the
``classical value" of $X$, while the second term is the
quantum correction. We see that it is the same for any $|n>$.
Any state $|\Psi>$ is a (normalized) linear combination of
$|n>$-states (like $|\alpha>$). So for any state
$|\Psi>$, we will get the same quantum correction.

The expectation value (14) describes a semiclassical dust
universe, starting with a maximum radius $a^{2}_{0}$, and
collapsing to a minimum radius\footnote{In the geometrical
units $M_{P} = l_{P} = \hbar^{1/2}$.}
   \be
a_{min} \sim \hbar = l_{P}
   \ee
here $\sim$ is because we should have the $p$-factor in (14).
This semiclassical universe {\em does not} reach the
classical singularity, $a=0$. The dust density, at
$a = a_{min}$ is
   \be
\rho_{max} \sim \frac{N}{a_{min}^{3}} = N l_{P}^{-3}
   \ee
where $N$ is the number of dust particles (``baryons").

According to our assumption that the metric is the
basic physical quantity, all other physical quantities
should be derived from $g_{\mu \nu}^{(\alpha)} =
<\alpha| \hat{g}_{\mu \nu}|\alpha>$. So we consider
$g_{\mu \nu}^{(\alpha)}$ as if it is a classical metric.
For example the corresponding Ricci scalar is
   \begin{eqnarray*}
R[g_{\mu \nu}^{(\alpha)}] = \frac{24}{ a_{0}^{2}(1 +
\mbox{cos}(t))^{2} + 2 l_{P}^{2} }
   \end{eqnarray*}
So the ``quantum Ricci scalar" is bounded from above,
$ R_{max} \sim (a_{min})^{-2} $.

In the case of the OS model, the solution (14) describes a
dust star that collapses to a minimum radius, $l_{P}$, and does
not reach the Schwarzschild singularity.

These results can be easily generalized to the (spherically
symmetric) inhomogeneous case.
Spherically symmetric dust matter in 4D Einstein gravity
is an integrable system [7].
One can fix the (diffeomorphism) gauge completely,
and get an infinite set of independent (non propagating)
degrees of freedom [1]. So we will first fix the gauge
classically, and then quantized the reduced system.
One can fix the gauge by choosing the comooving coordinates
in which the metric has the form
   \be
ds^{2} = -d\tau^{2} + e^{2\mu}d {\cal R}^{2} + e^{2\lambda}d\Omega_{2}^{2}
   \ee
where $\mu$ and $\lambda$ are functions of $\tau$ and ${\cal R}$, and
$d\Omega_{2}^{2}$ is the volume element in $S^{2}$. Explicitly, we choose
$\tau$ and ${\cal R}$ such that
   \begin{eqnarray*}
U_{\nu} &=& - \nabla_{\nu} \tau \\
{\cal R} &=& \lambda' e^{\lambda - \mu}
   \end{eqnarray*}
where $U_{\nu}$ are the 4-velocities of the dust particles,
and prime denote ${\cal R}$-differentiation. This gauge
fixing corresponds to $N_{\perp}=1$ , $N^{i}=0$ , and it is complete [1].
One then gets the reduced Hamiltonian [1,3]
   \be
H_{red} = \int_{0}^{\rho_{s}} \left( \frac{32\sqrt{1-\rho^{2}}}{3\pi}
\frac{P_{Y}^{2}}{Y} +
\frac{3\pi \rho}{2\sqrt{1 - \rho^{2}}} Y \right) d\rho
   \ee
where $\rho = \sqrt{1 - {\cal R}^{2}}$ , and $\rho_{s}$ is the
surface of the dust star.
$Y(\tau,\rho)$ is the (reduced) field variable, $ Y= 8 e^{\lambda} $ and
$P_{Y} = \delta L / \delta \dot{Y} = 3\pi Y \dot{Y} / (64 \rho
\sqrt{1 - \rho^{2}}) $. A most important feature of (19) is that there
are no $\rho$-derivatives in it. This means that the (infinite number
of) degrees of freedom are decoupled. One can descretize (19). Define
   \be
\rho_{k} = \frac{k}{N} \rho_{s} ~~~,~~ k = 1,2,...,N
   \ee
and write (19) as a sum
   \be
H_{red} = \frac{1}{N} \sum_{k=1}^{N} \left( \alpha_{k}
\frac{P_{k}^{2}}{Y_{k}} + \beta_{k} Y_{k} \right)
   \ee
where $Y_{k}(\tau)=Y(\tau,\rho_{k})$, $P_{k}(\tau)=P(\tau,\rho_{k})$,
and
   \begin{eqnarray*}
\alpha_{k} = \frac{32}{3\pi} \rho_{k} \sqrt{1-\rho_{k}^{2}}
{}~~~,~~~ \beta_{k} = \frac{3\pi}{2} \frac{\rho_{k}}{\sqrt{1 -
\rho^{2}_{k}}}
   \end{eqnarray*}
The classical equations of motion are $\dot{Y} = \delta H_{red} /
\delta P_{Y}$ and $\dot{P}_{Y}=-\delta H_{red} / \delta Y$. Given the
Cauchy data, $Y(\rho,\tau=0)~,~P_{Y}(\rho,\tau=0)$ (or
$\{ Y_{k}^{(0)} , P_{k}^{(0)} \}$), we get a unique
solution to the classical equations [7].

The Hamiltonian (19) (or (21)) is time independent so
we can write it as
   \be
H_{red} = \sum_{k=1}^{N} E_{k}
   \ee
where
   \be
E_{k}= \frac{1}{N} \left( \alpha_{k}
\frac{{(P_{k}^{(0)})}^{2}}{Y_{k}^{(0)}} + \beta_{k} Y_{k}^{(0)}
\right)
   \ee
and $Y_{k}^{(0)} = Y_{k}(\tau=0)$, $P_{k}^{(0)} = P_{k}(\tau=0)$.

To quantize this system, we use the coordinates representation
   \be
\hat{Y} = Y ~~~~~~;~~~~~~\hat{P}_{Y} = -i \hbar
\frac{\delta}{\delta Y}
   \ee
and get the Schr\"{o}dinger equation (the Wheeler-DeWitt equation)
   \be
i\hbar \frac{\partial \Psi[Y;t]}{\partial t} =
\int_{0}^{\rho_{s}} \left( - \frac{32 \sqrt{1-\rho_{k}^{2}}}{
3\pi} \frac{\hbar^{2}}{Y} \frac{\delta^{2}}{\delta Y^{2}} +
\frac{3\pi \rho}{2 \sqrt{1 - \rho^{2}}} Y \right) \Psi[Y;t] d\rho
   \ee
where we used the ``$YP_{Y}$ ordering" (the field $Y$ is
always to the left of its conjugate momenta). One can use
the discrete form of (25)
   \be
i\hbar \frac{\partial \Psi(\vec{Y},t)}{\partial t} =
\frac{1}{N} \sum_{k=1}^{N} \left( - \alpha_{k} \frac{\hbar^{2}}{Y_{k}}
\frac{\partial^{2}}{\partial Y_{k}^{2}} + \beta_{k} Y_{k}
\right) \Psi(\vec{Y},t)
   \ee
Using the fact that we have a decoupling, we can write
the quantum state, as a direct product
   \be
|\Psi> = |\Psi_{1}>|\Psi_{2}> \cdot \cdot \cdot |\Psi_{N}>
   \ee
and from (22), (25) and (26) we get
a set of $N$ independent (Schrodinger) equations
   \be
\frac{1}{N} \left( - \alpha_{k} \frac{\hbar^{2}}{Y_{k}}
\frac{\partial^{2}}{\partial Y_{k}^{2}} + \beta_{k} Y_{k}
\right) \Psi_{k}(Y_{k}) = E_{k} \Psi_{k}(Y_{k})
   \ee
which is the equation for an harmonic oscillator [3], with mass,
$m_{k}=1/2\alpha_{k}$, frequency, $\omega_{k}= 8 \rho_{k}$,
and energy, $\epsilon_{k} = E_{k}^{2} / 4 N^{2} \beta_{k} $.
So the spectrum of (26) is quantized, and we
can write $|\Psi_{k}>=|n_{k}>$. The Hilbert space of
solutions of the Schr\"{o}dinger equation (26) is spanned by
   \be
\{ |\Psi> \} = \{ |n_{1}>|n_{2}>\cdot \cdot \cdot |n_{N}> \}
   \ee
The quantization conditions, $\epsilon_{k} = \hbar \omega_{k}
(n_{k} + 1/2)$, give
   \begin{eqnarray*}
E_{k}^{2} = \frac{1}{N} \frac{ 12 \pi \hbar \rho_{k}^{2}}{\sqrt{
1 - \rho_{k}^{2}}} (n_{k} + 1/2)
   \end{eqnarray*}
and using (22) we get
   \be
H_{red} = \hbar^{1/2} \frac{1}{N} \sum_{k} A_{k} (n_{k}+1/2)^{1/2}
   \ee
where $ A_{k} = 4 \sqrt{3\pi} \rho_{k} / (1 - \rho_{k}^{2})^{1/4} $.
For a finite $N$, the spectrum (30) is discrete.

In the case of a dust star that collapses to form a
Schwarzschild black hole, $H_{red}$ is proportional to
the mass of the star. In that case (30) is the
``generalized Bekenstein area quantization" [8,3].

Because $A_{k}$ is of order $1$, an average $n_{k}$ is
of order $H^{2}_{red} / \hbar$, which for astronomical
objects is a huge number\footnote{For our sun for example,
we have $M_{\odot}^{2} / \hbar \sim 10^{60}$.}.
So we are in the semiclassical regime.

Consider a semiclassical collapse, starting
at rest (at $\tau=0$). In that case we have $P_{k}^{(0)}=0$,
and (28) is now
   \be
\left[ -\alpha_{k}\frac{\hbar}{Y_{k}}
\frac{\partial^{2}}{\partial Y_{k}^{2}} +
\beta_{k}(Y_{k} - Y_{k}^{(0)}) \right] \Psi_{k}(Y_{k}) = 0
   \ee
Defining $X_{k} \equiv Y_{k} - \frac{1}{2} Y_{k}^{(0)}$,
we see that $\{ X_{k} , P_{k} \}$ is a one dimensional
harmonic oscillator. A quasi-classical (coherent) state,
corresponding to a classical solution $X_{k}^{(cl)}$, is
   \be
\hat{a}_{_{X_{k}}}|\alpha_{k}> = \frac{X_{k}^{(0)}}{\sqrt{2}}
e^{-i \omega_{k} \tau} |\alpha_{k}>
   \ee

We want to calculate the metric expectation values. These
are $<\Psi| e^{2\mu} |\Psi>$ and $<\Psi| e^{ 2\lambda} |\Psi>$
(see (18)). In our gauge however, we have $e^{\mu} =
\frac{\sqrt{1 - \rho^{2}}}{8\rho} \partial Y / \partial \rho$.
So we need to calculate only $<\alpha_{k}| Y_{k}^{2} |\alpha_{k}>$.
Again, this is elementary, we use the $p$-ordering and get
   \be
<\alpha_{k}|Y_{k}^{2}|\alpha_{k}> =
{\left( Y_{k}^{(cl)} \right)}^{2} + \frac{4\hbar}{3\pi}
p \sqrt{1 - \rho_{k}^{2}}
   \ee
where $Y_{k}^{(cl)}$ is the classical collapsing
solution. Like in the homogeneous case,
$<\alpha_{k}|Y_{k}^{2}|\alpha_{k}>$ {\em does not} reach the
classical singularity, $<Y^{2}>=0$. For a general quantum state (29),
we have the same quantum correction, so the expectation value is
generally everywhere regular.

\vspace{0.5cm}
In this work we use a week condition for a ``quantum avoidance"
of classical singularities:
given the quantum state of the system, the quantum geometry
is singular, only if its metric expectation value is.

In the case of spherically symmetric dust universes (or
stars), the expectation value $<\Psi| \hat{g}_{\mu \nu}
|\Psi>$ is everywhere regular. Semiclassically it describes
a dust universe (or a dust star) that collapse to a minimum
radius, the Planck radius, and {\em does not} reach the
classical singularity.

The existence of a singularity in the space-time structure
is strongly related to the ``information puzzle" during
the Hawking evaporation. The standard view is that if the
information is lost, it is lost ``inside" the singularity.
It may be interesting to study the Hawking radiation
with a ``semiclassical background" [9], $<\Psi| \hat{g}_{\mu \nu}
|\Psi>$.

\vspace{1cm}
{\bf Acknowledgment} \\
I would like to thank Stanley Deser for very helpful
discussions.

\newpage
{\bf REFERENCES}
\begin{enumerate}
\item F. Lund, {\em Phys.Rev.}{\bf D8}, 3253; 4229 (1973) \\
B.S. DeWitt, {\em Phys.Rev.}{\bf 160},1113, (1967) \\
J.A. Wheeler, in {\em Battelle Recontres}, eds. C.M. DeWitt
and J.A. Wheeler (NY, Weily, 1968)
\item J.H. Kung, ``Quantization of the Closed Minisuperspace
Models as Bound states", Harvard Univ. report, hepth/9302016
(1993)
\item Y. Peleg, ``The Wave Function of a Collapsing Star",
Brandeis Univ. Report No. BRX-TH-342, hepth/9303169 (1993);
``Quantum Dust Black Holes", Brandeis Univ. Report No.
BRX-TH-350, hepth/9307057 (1993)
\item R. Arnowitt, S. Deser and C.W. Misner, {\em Ann.Phys.}
(NY) {\bf 33}, 88 (1965); ``The Dynamics of General Relativity",
in {\em Gravitation: an Introduction to Current Research},
ed. L. Witten (Wiley, NY, 1962) \\
B.F. Schutz, Jr. {\em Phys.Rev.}{\bf D4}, 3559 (1971)
\item C.W. Misner, K.S. Thorne and J.A. Wheeler,
{\em Gravitation} (Freeman, San Francisco, 1973)
\item J.R. Oppenheimer and H. Snyder, {\em Phys. Rev.}
{\bf 56}, 455 (1939)
\item R.C. Tolman, {\em Pric.Nat.Acad.Sci.}{\bf 20},
169 (1934) \\
L.D. Landau and L.M. Lifshitz, {\em The Classical Theory
of Fields}, (Pergamon, Oxford, 1962)
\item J.D. Bekenstein, {\em Lett.Nuov.Cimen.}{\bf 11},
476 (1974)
\item C.R. Stephnes, G. 't Hooft and B.F. Whiting, ``Black
Hole Evaporation without Information Loss", Univ. of Utrecht
report no. THU-93/20, hepth/9310006 (1993)

\end{enumerate}

\end{document}